%% file: main.tex
\documentclass[a4paper,12pt]{article}
\usepackage[margin=25mm]{geometry}
\usepackage{amsmath}
\usepackage{amsfonts}
\usepackage{amssymb}
\usepackage{graphicx}
\usepackage{verbatim}
\usepackage{natbib}
\usepackage{hyperref}

\providecommand{\keywords}[1]
{
  \small	
  \textbf{\textit{Keywords---}} #1
}

\title{Startup Financing: Token vs Equity}
\author{Guangye Cao$^{1}$ \\
        \small $^{1}$University of Michigan \\
}
\date{August 2023} % Comment this line to show today's date

\begin{document}
\maketitle

\begin{abstract}
Why would a blockchain-based startup and its venture capital investors choose to finance by issuing tokens instead of equity? What would be their rates of return for each asset? This paper focuses on the liquidity difference between the two fundraising methods. Early-stage startup equity is illiquid due to a lack of secondary market, so investors need to hold the equity until the startup goes public or is acquired. On the other hand, investors' tokens become vested in four or less years, and can be traded on crypto exchanges once the startup's product is launched. I build a three-period model of an entrepreneur, two types of investors, and users. Some investors have unforeseen liquidity needs in the middle period that can only be met with tokens. I find that the entrepreneur obtains higher payoff by issuing tokens instead of equity, and the payoff difference increases with investors' probability of needing liquidity in the middle period, the depth of the token's market, and investors' risk aversion. Investors willingly accept a lower return for tokens due to the asset's earlier liquidity. 
\end{abstract} \hspace{10pt}

%TC:ignore

%TC:endignore

\newpage

\section{Introduction}

Initial coin offerings (ICO) emerged as a new method of early-stage startup financing in 2013 and skyrocketed in 2017-2018. $\$11.4$ billion worth of tokens were sold in 2018, compared to seed-round venture capital issuance of $\$14.9$ billion (\cite{cointelegraph}, \cite{crunchbase}). The craze subsided in 2019, when only $\$2.6$ billion was raised in the first half of the year. In an ICO, the entrepreneurs of blockchain-based startups sell cryptocurrency in exchange for bitcoin or ether\footnote{ETH or ether is the native token of the Ethereum blockchain}, which they can then sell for fiat money to finance the development of their product. The cryptocurrency can have a variety of uses. It can be spent to purchase the product of the startup or the product sold on its platform, in which case it is called a utility token. In decentralized finance startups, the utility token can be deposited as collateral to mint new assets like stablecoins. Cryptocurrency can also represent a share of the profit of the startup, in which case it is called a security token. 

This paper focuses on the liquidity differences between tokens and equity to answer two questions: i) when would a blockchain-based startup and venture capital investors choose to finance with tokens instead of equity; ii) what would be their rates of return for each asset?

 I build a three-period model that compares equity and tokens as means of financing for an early stage startup building a blockchain-based product. Equity of startups is illiquid due to the lack of a secondary market, whereas tokens can be traded on cryptocurrency exchanges shortly after issuance. The model has an entrepreneur, investors (venture capital), and users. Investors have liquidity needs in the middle period that can only be met with tokens. The entrepreneur's preference for tokens increases with token liquidity and investors' probability of needing liquidity in the middle period. Investors willingly accept a lower return in exchange for higher liquidity. 

\section{Literature Review}

Much of the ICO literature centers around network-externality advantages of tokens for digital platform adoption, where the initial investors are also future users of the platform. \cite{xiongsockin} investigate how tokenization resolves the conflict between platforms and users. By giving governance rights to users, tokenization prevents platforms from exploiting users, which comes at the cost of not having an owner with equity stake who would subsidize participation to maximize network effects. \cite{congliwang1} show that token price reflects agents' expectation of future popularity of the platform, as a permanent positive shock to platform productivity not only increases user base today, but also increases expectation of future user base and future demand for tokens, which leads agents to invest in tokens today. Other works related to network effects include \cite{bakoshalaburda}, \cite{gryglewicz}, and \cite{limann}. Another branch of ICO papers focuses on moral hazard and agency problems. For example, \cite{chodlyandres}, and \cite{gantsoukalasnetessine} compare token and equity financing and argue that, token financing introduces a new agency problem - the entrepreneur underproduces the product or service, because he incurs all of the costs of production but is only entitled to a portion of the revenue. \cite{malinovapark} argue that a variation of traditional ICO mechanism offers stronger incentives for the entrepreneur to exert effort.

To the best of my knowledge, mine is the first paper to abstract from the potentially transient differences caused by a lack of regulation, to shed light on a key difference between the two, liquidity. As ICOs have moved away from crowdfunding back towards venture capital, much of the moral hazards can be mitigated by the staged VC financing process. I therefore model startup financing where the investors are accredited investors instead of future customers. Lastly, for investors, tokens are reminiscent of the demand deposits of \cite{diamonddybvig} and \cite{prescottwallace}. One difference is that the second period withdrawal (equivalent to the addiitional issuance of tokens) does not depend on the proportion of depositors who have withdrawn in the first period.

On the empirical side, \cite{fahlenbrachfrattaroli} show that many ICO investors resell their tokens on the secondary market. \cite{howellniessneryermack} studies the determinants of ICO success and found positive correlation with the amount of information disclosed to investors. Other notable papers include, and are not limited to: \cite{lyandrespalazzorabetti}, \cite{catalinigans}, and \cite{goldsteinguptasverchkov}.

\section{Model}

Consider a simple model in which tokens have two functions: i) as the medium of exchange to purchase the goods / services produced by the issuer (the entrepreneur), ii) as a means to raise funding for the issuer.  The entrepreneur needs funding from venture capitalists to build a blockchain or a decentralized application (dApp) that will produce a digital good / service. The entrepreneur can issue either tokens or equity, where the key difference between the two assets is their liquidity. As there does not exist an exchange for trading private equity of early-stage startups, VCs normally cannot sell the equity for five to ten years, until the startup is acquired or goes public. On the other hand, tokens could be listed on public cryptocurrency exchanges within months after the private issuance. Tokens are not perfectly liquid, however, as newly listed tokens start with low trading volume due to their obscurity. Additionally, VCs are often constrained by a vesting period, so they cannot sell all the tokens at once. The model captures the liquidity advantage of tokens (relative to equity) to show that the initial price of tokens increases with liquidity. Furthermore, there is a threshold level of liquidity above which the entrepreneur can earn higher profit by selling tokens instead of selling equity.

\subsection{Setup}

 There are three players: an entrepreneur, VC investors, and users; two goods: a digital good and a generic good (numeraire); three time periods: $t = \{0,1,2\}$, and three assets: a risk-free bond, tokens, and equity. I use the notation $'$ for the entrepreneur, $''$ for the users, and no subscript for the investors. Assume that the initial investment $I$ produces an exogenous stream of digital good $\{y'_1, y'_2\}$, which can be interpreted as the maximum capacity or throughput of the dApp. To finance $I$, the cost of developing the dApp, the entrepreneur can issue equity or tokens to investors. Equity pays a dividend only in $t=2$ and cannot be traded at all in $t=1$. This can be interpreted as the startup needs to maintain a cash reserve by keeping the retained earnings instead of distributing them to the shareholders. Tokens can be traded in both $t=1$ and $t=2$; however, only a portion $\{\phi_1, \phi_2\}$ of investors' tokens can be sold in each period. One can interpret this exogenous liquidity parameter $\phi_t$ as the portion of tokens that can be sold with zero transaction fee, while the remaining $1-\phi_t$ has infinite transaction fee. Alternatively,if $\phi_1>0$ and $\phi_2=1$, the parameter represents the share of investor's tokens that has become vested each period.

There are two types of investors: type $a$ consumes only in $t=1$, and type $b$ consumes only in $t=2$. In $t=1$, investors learn of their own types. In $t=0$, they only know that the probabilities of being types $a$ and $b$ are $\lambda$ and $1-\lambda$, respectively. One can think of this "consumption need" as an unexpected investment opportunity. In contrast, the entrepreneur consumes only in $t=2$ and users consume in both $t=1$ and $t=2$. I further assume that users can buy both digital and generic goods, while the other players can buy only the generic good. Lastly, investors and the entrepreneur have access to all three assets, but users can only save in bonds.

The purpose of these assumptions is to isolate the effect of liquidity on token and equity pricing while capturing token's function as a medium of exchange. The assumptions that only users can buy the digital good, but they cannot save in tokens, cleanly separate the roles of users and investors. Users would not face a trade-off between keeping or selling tokens for capital gain and spending them for consumption. The assumption that the entrepreneur consumes only in $t=2$ makes equity a feasible financing instrument for him despite its illiquidity. 

I further assume that the user has perfect-substitute utility function. This means that the price of the digital good would be equal to one unit of the generic good regardless of the entrepreneur's financing choice. Lastly, when the digital and generic goods cost the same, he will consume the digital good first. In other words, the user will buy all the digital good that is produced, within his budget constraint.

The timing is as follows. At $t=0$, the entrepreneur either i) sells $T_0$ tokens at price $p_0$ ($p_0 T_0 = I$), or ii) create 1 share of equity priced at $q$ and sells a fraction $e$ of the share to investors ($eq=I$). Investors have initial wealth $W$, which they allocate between equity and bonds, or tokens and bond, depending on what the entrepreneur issued.  The risk-free bond is completely liquid and can be traded each period costlessly. Users invest all their wealth ($W''$) into bonds.

At the beginning of $t=1$, the entrepreneur launches the completed dApp to the public, and he incurs fixed cost $\omega$  to produce $y'_1$ units of digital good. Tokens become tradeable on a cryptocurrency exchange, and investors learn of their own types. Next, asset markets clear. The entrepreneur can issue additional tokens $T'_1$ to users and investors on the exchange at price $p_1$, and investors can sell up to $\phi_1 T_0$ of their previous token holding. They can also purchase bonds. Finally, users can redeem their tokens immediately for the digital good, at one token per good.  If equity was issued instead, users would buy digital good with fiat money instead of tokens. Assuming the asset market clears before the goods market prevents the entrepreneur from buying back equity, as startups would generally re-invest any retained earnings back into the business. 

At $t=2$, the same process repeats. If equity was issued, the accumulated dividend would be distributed. Type $b$ investors and entrepreneur consume the generic good. 

The only variables that need to be solved for are the price and quantity of tokens and equity issued in $t=0$ and entrepreneur's payoff from each financing method, $\{p_0, T_0, e, q, c^{'E}_{2}, c^{'T}_{2} \}$. All other variables are exogenous parameters. In the next two subsections, I solve the model with risk-neutral and risk-averse investors, separately. I will characterize the price and the required rate of return of each asset, and the entrepreneur's payoff. 

\subsection{Risk-neutral investors}

\subsubsection{Equity}
\textbf{User}
The user starts in period $t=0$ with endowment $W''$. Let $B_t''$, $c_t''$, and $y_t''$ be his risk-free bond holding, generic good, and digital good consumption, respectively. Since he can only save in bonds, $W''=B_0''$. His optimization problems for $t\in \{1,2 \}$ are:

\begin{align}
    \max_{c''_1, y''_1, B''_1} (c''_1 + y''_1) + \beta \mathbb{E}_1 (c''_2 + y''_2) \quad & \text{subject to:} \quad B''_0 R = c''_1 + y''_1 p_1 + B''_1\\ \nonumber
    \max_{c''_2 , y''_2} c''_2 + y''_2 \quad & \text{subject to:} \quad  B''_1 R = c''_2 + y''_2 p_2
\end{align}

\textbf{Investors} At $t=0$, before knowing their own types, the investors allocate their wealth between bonds and equity to maximize expected future utility. Let $\Pi$ denote the future value of profit, $\Pi = (y_1'-\omega)R + y_2' - \omega$.

\begin{equation}
    \max_{B_0, e} \lambda \beta \mathbb{E}_0  (\underbrace{B_0 R}_{c_1}) + (1-\lambda) \beta^2 \mathbb{E}_0(\underbrace{B_0 R^2 + e \Pi}_{c_2}) \quad \text{subject to} \quad W = B_0 + eq
\end{equation}

At $t=1$, type $a$ investors sell all of their bonds to consume. At $t=2$, type $b$ investors consume their share of the startup's profit and the future value of bonds. Taking the derivative w.r.t. $B_0$, $e$, and applying $\beta = \frac{1}{R}$, the price of equity ($q$) is equal to the present value of expected future payoff. $\beta = \frac{1}{R}$ ensures that the players are indifferent between consuming today or saving for tomorrow.

\begin{equation}
    q = \frac{(1-\lambda)\Pi}{R^2}
\end{equation}

With $\lambda >0$, the investors expects liquidity need in $t=1$. While bonds can be sold to meet that need, equity cannot. To compensate,  eq(\ref{eq: R^E_n}) shows that equity's required gross rate of return, $R^E$, would be greater than the risk-free rate. Investors' return is $IR^E$.

\begin{equation}
    R^E = \frac{\Pi}{q} = \frac{R^2}{1-\lambda} \label{eq: R^E_n}
\end{equation}

In $t=1$, type $a$ investors try to sell their equity to outside interests, but can only do so at a steep discount because there does not exist a liquid market for startup equity. The assumption here is that the discount price is 0, which means that type $a$ investors' share of profits, $\lambda e \Pi$, goes to the outside interests. 

\noindent \textbf{Entrepreneur}
The entrepreneur will receive his share of accumulated profit at $t=2$ and spend it all on the generic good ($c'_2$). The retained earnings from $t=1$ accrues gross risk-free interest $R$. Equation \ref{eq: c^E_n} shows that the entrepreneur's payoff is equal to the future value of profit subtracted by investors' return.

\begin{equation}
    c^{'E}_{2,n} = (1-e) \Pi = \Pi - \frac{I R^2}{1-\lambda} = \Pi - I R^E \label{eq: c^E_n}
\end{equation}

\subsubsection{Tokens}
\textbf{Investors}
\begin{align*}
    \max_{B_0, T_{0,n}} \lambda \beta \mathbb{E}_0 \underbrace{(B_0 R + \phi_1 T_{0,n}}_\text{$c_1$}) & + (1-\lambda)\beta^2  \mathbb{E}_0 [\underbrace{(B_0 R + \phi_1 T_{0,n})R + \phi_2(1-\phi_1)T_{0,n}}_\text{$c_2$}] \\
    \text{subject to:}\quad & W = B_0 + T_{0,n} p_{0,n} \quad \text{and}\\
            & T_{1,n} \geq (1-\phi_{0})T_{0,n} \quad \text{(liquidity constraint)}
\end{align*}

Because $p_1 = p_2 = 1$, investors will not earn returns from holding tokens from $t=1$ to $t=2$. They will sell as much of their token holding as they can, so the liquidity constraint is binding $T_{1,n} = (1-\phi_1)T_{0,n}$. Taking the derivative w.r.t. $B_0$ and $T_{0,n}$, and applying $\beta = \frac{1}{R}$, eq. \ref{eq: p_n} shows that the price of token ($p_{0,n}$) is equal to the present value of the fraction of tokens sold in $t=1$, plus the present value of the additional portion sold in $t=2$ multiplied by the probability of needing liquidity in $t=2$. Since some tokens cannot be sold in $t=1$ to meet the needs of type $a$ investors, tokens become less valuable as the need for early consumption ($\lambda$) increases. Token price increases in $\phi_1$ and $\phi_2$ as being able to offload tokens in each period makes the asset more valuable. 

\begin{equation}
    p_{0,n} = \frac{\phi_1}{R} + \frac{(1-\lambda)\phi_2(1-\phi_1)}{R^2}\label{eq: p_n}
\end{equation}

In eq. \ref{eq: p_n}, $\frac{\phi_1}{R}$ is the present value of utility from selling tokens in $t=1$, and 
$\frac{(1-\lambda)\phi_2(1-\phi_1)}{R^2}$ is the utility from selling additional tokens in $t=2$ if the investor is type $b$. The expected required gross rate of return by selling at $t=1$ for $p_1 = 1$ is:

\begin{equation}
    R^T_n = \frac{1}{p_{0,n}} = \frac{R^2}{\phi_1 R + (1-\lambda)\phi_2(1-\phi_1)}
\end{equation}

\noindent \textbf{Entrepreneur}: The entrepreneur's payoff is the value of $t\in\{1,2\}$ token issuance net of $\omega$, which depends on the parameters $\phi_1$ and $\phi_2$. The tokens he issues in a period is equal to digital good output minus the quantity of tokens resold by investors; $T'_{1,n} = y'_1 - \phi_1 T_{0,n}$ and $T'_{2,n} = y'_2 - \phi_2(1-\phi_1)T_{0,n}$. Equation \ref{eq: c_n} shows that entrepreneur's payoff decreases in $\lambda$. Note that $T_{0,n} = \frac{IR^2}{\phi_1 R + (1-\lambda)\phi_2(1-\phi_1)}$ is increasing in $\lambda$, meaning as the probability of needing early liquidity increases, token price must decrease to compensate for the asset's imperfect liquidity in $t=1$. As more tokens must be issued in $t=0$ to finance $I$, fewer will be issued by the entrepreneur afterwards.

\begin{align}
    \nonumber c^{'T}_{2,n} &= T_{2,n}' p_2 + (T'_{1,n} p_1 - \omega)R -\omega    \\
     \nonumber &=(y_1 - \omega)R + y_2 - \omega - IR^2(\frac{1}{\frac{\lambda R \phi_1}{\phi_2(1-\phi_1)+\phi_1 R} +1-\lambda})\\
    &=\Pi - \frac{IR^2}{\frac{\lambda R \phi_1}{\phi_2(1-\phi_1)+\phi_1 R} +1-\lambda}\label{eq: c_n} \\
    \nonumber &\geq c^{'E}_{2,n} = \Pi - \frac{IR^2}{1-\lambda} 
\end{align}

\noindent \textbf{Result 1}: $\frac{\partial c^{',T}_{2,n}}{\partial \phi_1}>0$ and $\frac{\partial c^{',T}_{2,n}}{\partial \phi_2}<0$. \textit{Proof}: see technical appendix. 

The entrepreneur's payoff increases in $\phi_1$ and decreases in $\phi_2$. Higher $\phi_t$ makes tokens more valuable so fewer are sold in $t=0$, which benefits the entrepreneur as he will be able to sell more new tokens in $t\in \{1,2\}$ to meet demand for digital good. On the other hand, higher $\phi_t$ also allows investors to resell more tokens in $t\in \{1,2\}$, so the entrepreneur would issue fewer new tokens. For $\phi_2$, the latter effect is stronger than the former. In the special cases where $\phi_1 = 0$ and $\phi_2 >0$ or $\phi_1 > 0$ and $\phi_2 = 0$, the number of tokens resold by investors, $\phi_1 T_0 = \phi_1 \frac{IR}{\phi_1}$ and $\phi_2 T_0 = \phi_2 \frac{IR^2}{(1-\lambda)\phi_2}$, do not depend on $\phi_t$, meaning that $T_0$ adjusts perfectly to keep entrepreneur payoff independent of $\phi_t$.\\

\noindent \textbf{Result 2}: $c^{'T}_{2,n}>c^{'E}_{2,n}$ if $\phi_1>0$ and $\lambda>0$. 

Equation \ref{eq: c_n} shows that if $\phi_1>0$ and $\lambda>0$, then $\frac{\lambda R \phi_1}{\phi_2(1-\phi_1)+\phi_1 R} +1-\lambda>1-\lambda$ $\Rightarrow$ $\frac{IR^2}{\frac{\lambda R \phi_1}{\phi_2(1-\phi_1)+\phi_1 R} +1-\lambda} < \frac{IR^2}{1-\lambda}$ $\Rightarrow$ $\Pi - \frac{IR^2}{\frac{\lambda R \phi_1}{\phi_2(1-\phi_1)+\phi_1 R} +1-\lambda}\label{eq: c_n} \geq \Pi - \frac{IR^2}{1-\lambda}$. This means as long as there is positive need for early liquidity and tokens can partially satisfy that need, the entrepreneur is better off issuing tokens instead of equity. Even if $\phi_2 = 0$, $c^{'T}_{2,n} = \Pi - IR^2 > c^{'E}_{2,n} = \Pi - \frac{IR^2}{1-\lambda}$. Intuitively, even if type $b$ investors cannot sell any tokens in $t=2$, they can sell some tokens in $t=1$ and still be able to consume in $t=2$. The value of $\phi_2$ does not matter because $\phi_1>0$ guarantees consumption in both period. \\

\noindent \textbf{Result 3}: Equity is the same as tokens that can be sold only in $t=2$, ie: $\phi_1=0$, $\phi_2=1$. In this case, $p_0 = (1-\lambda)\beta^2 = \frac{1-\lambda}{R^2}$. Investors' required rate of return and entrepreneur's payoff are the same as those with equity: $R^T = \frac{1}{p_0} = \frac{R^2}{1-\lambda}=R^E$ and $c^{'T}_{2,n} = \Pi  - \frac{IR^2_f}{1-\lambda} = c^{'E}_{2,n}$. \\

\noindent \textbf{Result 4}: The risk-free bond is the same as tokens can be all sold in $t=1$. In this case, $p_0 = \frac{1}{R}$, $R^T = R$, and $c^{'T}_{2,n} = \Pi - IR^2$.

To summarize, with risk-neutral investors, the price of tokens at $t=0$ is the present value of utility from selling tokens in $t=1$ plus the expected utility from selling additional token in $t=2$ if the investor is revealed to be type $b$. As long as some tokens can be sold in $t=1$, ie. $\phi_1>0$, tokens guarantee that both types of consumers will be able to consume, while equity allow only type $b$ consumers to consume. 

\subsection{Risk-averse investors}
\subsubsection{Equity Financing}

\textbf{Investors}: 
With constant relative risk aversion, investors' desire to smooth consumption increases with curvature ($\sigma$) of the utility function. Equity cannot satisfy this desire at all as it gives investors no return at $t=1$. Hence, CRRA investors value equity less than risk-neutral investors would (equation \ref{eq: q_a}).
\begin{equation}
    \max_{B_0, e_a} \lambda \beta \mathbb{E}_0 \frac{(\overbrace{B_0 R}^{c_1})^{1-\sigma}-1}{1-\sigma} + (1-\lambda)\beta^2  \mathbb{E}_0 \frac{(\overbrace{B_0 R^2 + e_a\Pi}^{c_2})^{1-\sigma}-1}{1-\sigma} \quad \text{subject to} \quad W = B_0 + e_a q
\end{equation}
\begin{align}
   q_a &= \frac{(1-\lambda)\Pi}{R^2\big[\lambda(\frac{B_0 R^2 + e_a\Pi}{B_0 R})^\sigma + 1-\lambda   \big]} \leq \frac{(1-\lambda)\Pi}{R^2} = q_n \label{eq: q_a}\\
   R^E_a &= \frac{\Pi}{q_a} = \frac{R^2}{1-\lambda}\Big[\lambda(\frac{B_0 R^2 + e_a\Pi}{B_0 R})^\sigma + 1-\lambda \Big] \geq \frac{R^2}{1-\lambda} = R^E_n \label{eq:R^E_a}
\end{align}
A larger $\frac{B_0 R^2 + e_a\Pi}{B_0 R} = \frac{c_2}{c_1}$ indicates less consumption smoothing. In equation \ref{eq:R^E_a} shows that investors require a risk premium $\lambda (\frac{B_0 R^2 + e_a\Pi}{B_0 R})^\sigma + 1-\lambda$ that increases in $\frac{c_2}{c_1}$, leaving less profit for the entrepreneur.

\noindent \textbf{Entrepreneur}
\begin{equation}
    c^{'\text{E}}_{2,a}  = \Pi - \frac{IR^2}{1-\lambda}\Big[\lambda(\frac{B_0 R^2 + e_a\Pi}{B_0 R})^\sigma + 1-\lambda \Big] \leq c^{'\text{E}}_{2,n}
\end{equation}

\subsubsection{Token Financing}
\textbf{Investor}

Equation \ref{eq: p_a} shows that token price with CRRA investors is equal token price with risk-neutral investors adjusted for the ratio $\frac{c_2^{-\sigma}}{\lambda c_1^{-\sigma} + (1-\lambda)c_2^{-\sigma}}$, which represents period 2 marginal utility as a share of expected marginal utility. A smaller ratio indicates less consumption smoothing with tokens.
\begin{align}
    p_{0,a} &= \frac{\phi_1}{R}+\frac{(1-\lambda)\phi_2(1-\phi_1)}{R^2}\frac{\overbrace{c_2^{-\sigma}}^{\text{MU}_2}}{\underbrace{\lambda c_1^{-\sigma} + (1-\lambda)c_2^{-\sigma}}_{\text{expected marginal utility}}} \label{eq: p_a}\\
    \nonumber    &\leq \frac{\phi_1}{R}+\frac{(1-\lambda)\phi_2(1-\phi_1)}{R^2} = p_{0,n}
\end{align}

\noindent \textbf{Entrepreneur}

\begin{figure}[h!]
    \centering
    \includegraphics[width = 4.5in]{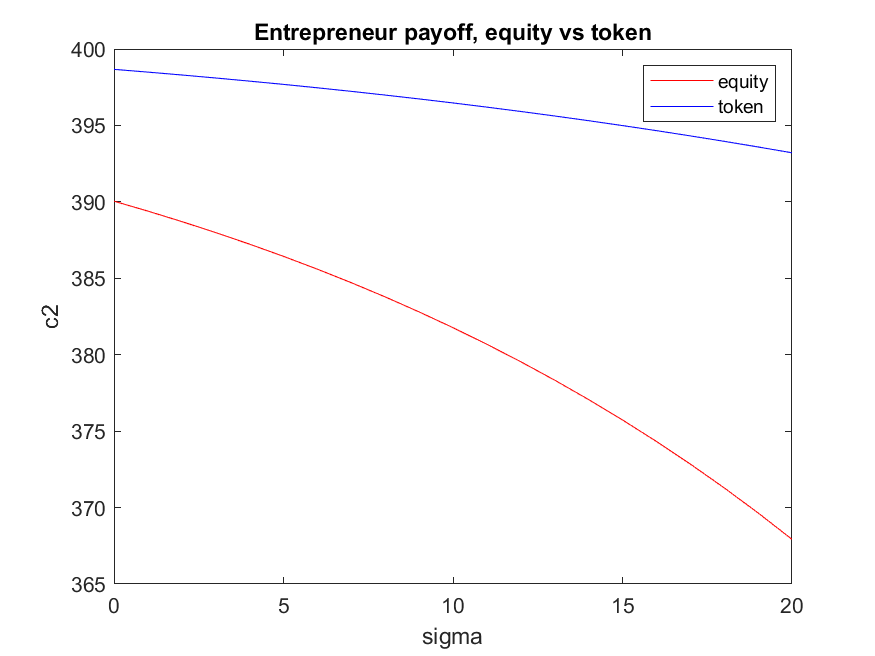}
    \caption{Entrepreneur Payoff, Equity vs. Token, for $\lambda=0.1$}
    \label{fig:c2_crra}
\end{figure}

\noindent \textbf{Result 5:} The entrepreneur's payoff (consumption in $t=2$) declines as investors become more risk-averse (as $\sigma$ increases). The decline is more steep if the entrepreneur finances with equity instead of tokens. See figure \ref{fig:c2_crra} as illustration.

\begin{align}
    c^{'\text{T}}_{2,a} &= \Pi-\frac{IR^2 (\lambda c_2^\sigma + (1-\lambda)c_1^\sigma)(\phi_2(1-\phi_1)+\phi_1 R)}{\phi_1 R (\lambda c_2^\sigma + (1-\lambda)c_1^\sigma)+(1-\lambda)\phi_2(1-\phi_1)c_1^{\sigma}}\\ \nonumber
     &\leq  \Pi - IR^2\frac{\phi_2(1-\phi_1)+\phi_1 R}{\phi_1 R + (1-\lambda)\phi_2(1-\phi_1)} = c^{'\text{T}}_{2,n}
\end{align}

 \section{Conclusion}
 This simple model compares an entrepreneur's payoffs when financing by issuing equity vs tokens to venture capital. Tokens allows both types of investors to consume, whereas equity satisfies only the investors with late consumption needs. Before knowing their own types, investors are willing to accept a lower rate of return in exchange for the liquidity benefits of tokens. For the entrepreneur, tokens give higher payoff than equity, and the difference in payoff increases with investors' risk aversion. 

\appendix 
\input{appendix}

\bibliographystyle{apalike}
\bibliography{reference}

\end{document}

%% file: appendix.tex
\section{Technical Appendix}
\subsection{Risk Neutral}
\subsubsection{Equity}

\begin{equation}
    \mathcal{L} = \lambda \beta (B_0 R) + (1-\lambda)\beta^2(B_0 R^2 + e[(y_1 - \omega)R + y_2 - \omega]) + \mu(W-B_0 - eq)
\end{equation}
\begin{align}
   \nonumber \partial B_0:& \quad \lambda\beta R + (1-\lambda)\beta^2 R^2 = \gamma \\
\nonumber   \partial e: & \quad (1-\lambda)\beta^2 [(y_1-\omega)R + y_2 - \omega ] = \gamma q \\
\nonumber \partial \mu: & \quad W = B_0 - eq\\
    q &= \frac{(1-\lambda)[(y_1-\omega)R + y_2 - \omega ]}{R^2}
\end{align}

Required rate of return: $R^E = \frac{(y_1 - \omega)R + y_2 - \omega}{q} = \frac{R^2}{1-\lambda}$
\begin{align}
    \nonumber c'_{2\text{equity}} &= (1-e) [(y_1-\omega)R + y_2 - \omega]\\
    &=(y_1 - \omega)R + y_2 - \omega - \frac{I R^2}{1-\lambda}
\end{align}

\subsubsection{Tokens}
    \begin{equation}
    \mathcal{L} = \lambda \beta (B_0 R_f + \phi_1 T_0) + (1-\lambda)\beta^2[(B_0 R_f + \phi_1 T_0)R_f + \phi_2(1-\phi_1)T_0] + \mu(W-B_0 - T_0 p_0)
\end{equation}
\begin{align*}
    \partial B_0:& \quad \lambda \beta R_f + (1-\lambda)\beta^2 R_f^2 = \mu\\
    \partial T_0:& \quad \lambda \beta \phi_1 + (1-\lambda)\beta^2 [\phi_1 R_f + \phi_2(1-\phi_1)] = \mu p_0\\
    \partial \mu: & \quad W=B_0 - T_0 p_0
\end{align*}
\begin{equation}
    p_0 = \frac{\phi_1}{R_f} + \frac{(1-\lambda)\phi_2(1-\phi_1)}{R_f^2}
\end{equation}
Gross required rate of return (sell at $t=1$ for 1):
\begin{equation}
    \mathbb{E}_0(R^T) = \frac{1}{p_0}
\end{equation}
\begin{align}
    \nonumber c'_{2\text{token}} &= T_2' p_2 + B_1' R_f -\omega   \\
    \nonumber  &= y_2 -\phi_2(1-\phi_1)T_0 + R_f(y_1 - \phi_1 T_0 - \omega) - \omega\\
    \nonumber       &=(y_1 - \omega)R_f + y_2 - \omega - (\phi_2(1-\phi_1)+\phi_1 R_f)\frac{I}{p_0}\\
     \nonumber       &=(y_1 - \omega)R_f + y_2 - \omega - (\phi_2(1-\phi_1)+\phi_1 R_f)IR^T\\
            &= (y_1 - \omega)R_f + y_2 - \omega - IR_f^2(\frac{\phi_2(1-\phi_1)+\phi_1 R_f}{\lambda R_f \phi_1 + (1-\lambda)(\phi_2(1-\phi_1)+\phi_1 R_f)}) \label{eq:neutral_c'2}
\end{align}

\textbf{Result 1}
\begin{align}
    \frac{\partial c^{',T}_{2,n}}{\partial \phi_1} &= IR^2\frac{R\lambda \phi_2}{[\phi_1(R - (1-\lambda)\phi_2)+(1-\lambda)\phi_2]^2}\geq 0 \\
    \frac{\partial c^{',T}_{2,n}}{\partial \phi_2} &= -IR^2 \frac{(1-\phi_1)\lambda \phi_1 R}{[\phi_1(R - (1-\lambda)\phi_2)+(1-\lambda)\phi_2]^2}    \leq 0
\end{align}

\subsection{Constant Relative Risk-Averse Investors}
\subsubsection{Equity}
\begin{equation*}
       \mathcal{L} = \lambda \beta \frac{(B_0 R_f)^{1-\sigma}-1}{1-\sigma} + (1-\lambda)\beta^2\frac{(B_0 R_f^2 + e\Pi)^{1-\sigma}-1}{1-\sigma} + \gamma(W-B_0 - eq)
\end{equation*}
\begin{align*}
    \partial B_0:& \quad \lambda \beta (B_0 R)^{-\sigma}R + (1-\lambda)\beta^2(B_0 R^2 + e\Pi)^{-\sigma} = \gamma\\
    \partial e:& \quad (1-\lambda)\beta^2(B_0 R^2 + e\Pi)^{-\sigma}\Pi = \gamma q \\
    \partial\gamma:& \quad W = B_0 + eq
\end{align*}
\begin{align*}
   q_a &= \frac{(1-\lambda)\beta^2(B_0 R + e\Pi)^{-\sigma}\Pi}{\lambda\beta(B_0 R)^{-\sigma}R + (1-\lambda)\beta^2(B_0 R + e\Pi)^{-\sigma}R^2} \\
   &=\frac{(1-\lambda)\Pi}{R^2\big[\lambda(\frac{B_0 R^2 + e\Pi}{B_0 R})^\sigma + 1-\lambda   \big]}
\end{align*}
\begin{equation*}
    R^E_a = \frac{\Pi}{q} = \frac{R^2}{1-\lambda}\Big[\lambda(\frac{B_0 R^2 + e\Pi}{B_0 R})^\sigma + 1-\lambda \Big]
\end{equation*}
\begin{align*}
    c^{'\text{E}}_{2,a} & = (1-e)\Pi = (1-\frac{I}{q})\Pi\\
        & = \Bigg[1 - \frac{IR^2\big[\lambda(\frac{B_0 R^2 + e\Pi}{B_0 R})^\sigma + 1-\lambda}{(1-\lambda)\Pi}    \Bigg]\Pi\\
        &= \Pi - \frac{IR^2}{1-\lambda}\Big[\lambda(\frac{B_0 R^2 + e\Pi}{B_0 R})^\sigma + 1-\lambda \Big]
\end{align*}

\subsubsection{Token}
\begin{align*}
    \mathcal{L} &= \lambda \beta \frac{(B_0 R_f + \phi_1 T_0)^{1-\sigma}-1}{1-\sigma}\\
    &+ (1-\lambda)\beta^2\frac{[(B_0 R_f + \phi_1 T_0)R_f + \phi_2(1-\phi_1)T_0]^{1-\sigma}-1}{1-\sigma} + \mu(W-B_0 - T_0 p_0)
\end{align*}
\begin{align*}
    \partial B_0: & \quad \lambda \beta (B_0 R + \phi_1 T_0)^{-\sigma}R + (1-\lambda)\beta^2 [(B_0 R + \phi_1 T_0)R + \phi_2(\phi_1)T_0]^{-\sigma}R^2 = \mu \\
    \partial T_0: & \quad \lambda \beta (B_0 R + \phi_1 T_0)^{-\sigma}\phi_1 + (1-\lambda)\beta^2 [(B_0 R + \phi_1 T_0)R + \phi_2(\phi_1)T_0]^{-\sigma}[\phi_1 R + \phi_2(1-\phi_1)]\\
     &= \mu p_0\\
    \partial \mu: & \quad W = B_0 + T_0 p_0
\end{align*}

Let $\Phi = \phi_1 R + \phi_2(1-\phi_1)$.
\begin{align*}
    p_{0,a} & = \frac{\lambda \beta (B_0 R + \phi_1 T_0)^{-\sigma}\phi_1 + (1-\lambda)\beta^2 [(B_0 R + \phi_1 T_0)R + \phi_2(1-\phi_1)T_0]^{-\sigma}\Phi}{\lambda \beta (B_0 R + \phi_1 T_0)^{-\sigma}R + (1-\lambda)\beta^2 [(B_0 R + \phi_1 T_0)R + \phi_2(1-\phi_1)T_0]^{-\sigma}R^2}\\
    & = \frac{\frac{\lambda \beta \phi_1}{c_1^{\sigma}} + \frac{(1-\lambda)\beta^2[\phi_1 R + \phi_2(1-\phi_1)]}{c_2^{\sigma}}}{\frac{\lambda \beta R}{c_1^{\sigma}} + \frac{(1-\lambda)\beta^2 R^2}{c_2^{\sigma}}}\\
    &=\frac{\beta \phi_1(\lambda c_2^{\sigma} + (1-\lambda)c_1^{\sigma}) + (1-\lambda)\beta^2\phi_2(1-\phi_1)c_1^{\sigma}}{\lambda c_2^{\sigma} + (1-\lambda)c_1^{\sigma}}\\
    &= \frac{\phi_1}{R}+\frac{(1-\lambda)\phi_2(1-\phi_1)}{R^2}\frac{c_1^\sigma}{\lambda c_2^\sigma + (1-\lambda)c_1^\sigma}
\end{align*}
\begin{align*}
    R_a^T &= \frac{1}{p_{0,a}}= \frac{R^2 (\lambda c_2^\sigma + (1-\lambda)c_1^\sigma)}{\phi_1 R [\lambda c_2^\sigma + (1-\lambda)c_1^\sigma]+(1-\lambda)\phi_2(1-\phi_1)c_1^{\sigma}}
\end{align*}
\begin{align*}
    c^{'\text{T}}_{2,a} &= \Pi - (\phi_2(1-\phi_1)+\phi_1 R)\frac{I}{p_0}\\
    &= \Pi-(\phi_2(1-\phi_1)+\phi_1 R)\frac{IR^2 (\lambda c_2^\sigma + (1-\lambda)c_1^\sigma)}{\phi_1 R (\lambda c_2^\sigma + (1-\lambda)c_1^\sigma)+(1-\lambda)\phi_2(1-\phi_1)c_1^{\sigma}}
\end{align*}